\documentclass[pdflatex,sn-mathphys]{sn-jnl}
\newcommand{\alf}{{Alfv{\'e}n }}

\graphicspath{{./}{Figures/}}
\jyear{2023}%

\begin{document}

\title[Formation of Magnetic Switchbacks]{ Formation of Magnetic Switchbacks Observed by Parker Solar Probe}


\author*[1]{\fnm{Gabor} \sur{Toth}}\email{gtoth@umich.edu}

\author[2]{\fnm{Marco} \sur{Velli}}\email{mvelli@ucla.edu}

\author[1]{\fnm{Bart}
\sur{van der Holst}}\email{bartvand@umich.edu}

\affil*[1]{\orgdiv{Department of Climate and Space Sciences and Engineering}, \orgname{University of Michigan}, \orgaddress{\street{2455 Hayward}, \city{Ann Arbor}, \postcode{48109}, \state{MI}, \country{USA}}}

\affil[2]{\orgdiv{Department of Earth, Planetary and Space Sciences}, \orgname{University of California at Los Angeles}, \orgaddress{\street{603 Charles E. Young Drive, East}, \city{Los Angeles}, \postcode{90095}, \state{CA}, \country{USA}}}


\maketitle



\begin{figure}
\includegraphics[width=\textwidth]{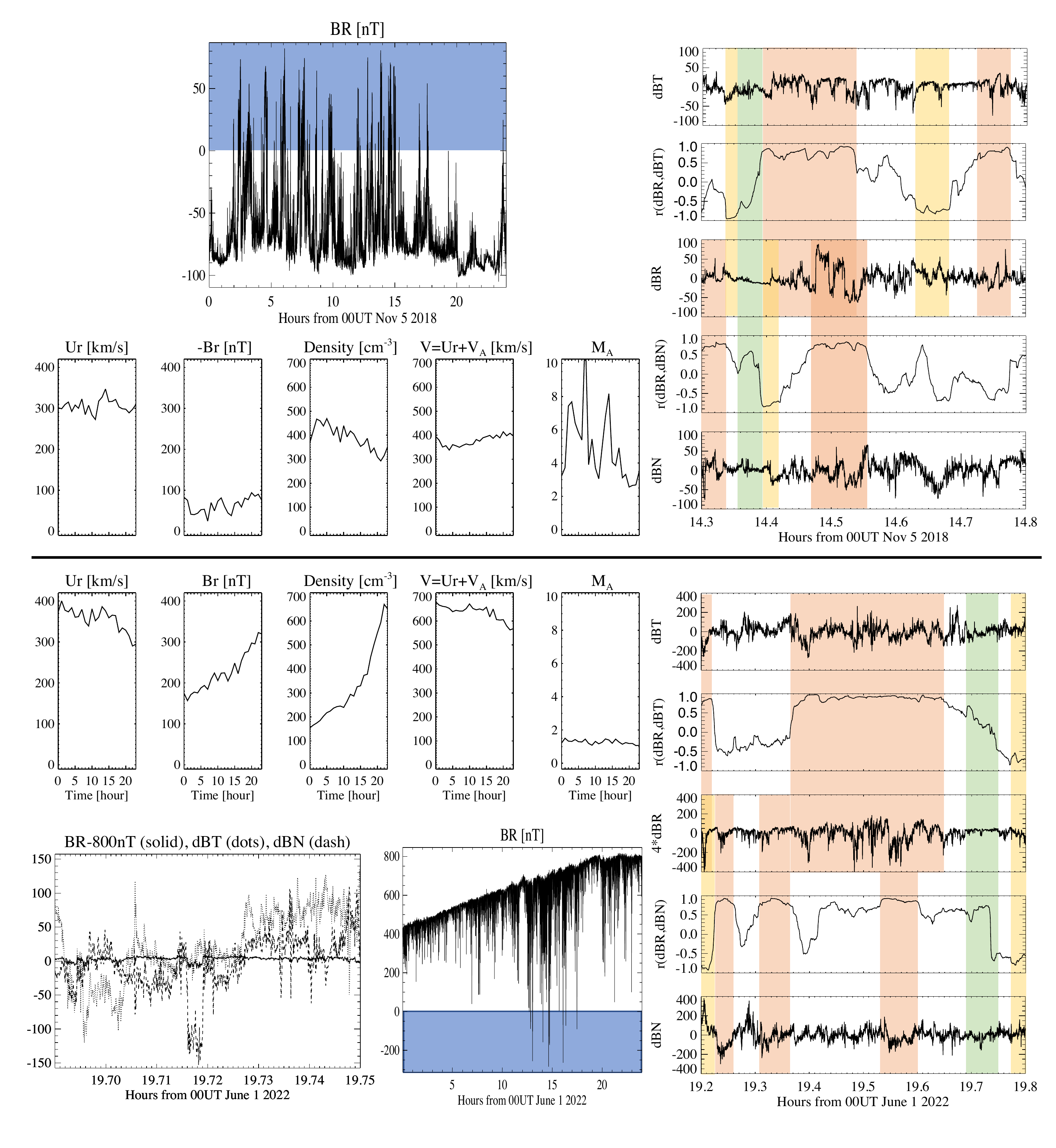}
\vspace{-0.75cm}
\caption{
Parker Solar Probe observations of switchbacks. The figures in the top half are from November 5 2018 during the first encounter at about $35\,R_s$ from the Sun, and the rest are from May 30 to June 1 2022 during the 12th encounter between $30\,R_s$ and $13.3\,R_s$ distance. The two panels showing the radial field $B_r$ with 0.22\,s time cadence contain switchbacks where the black line is inside the blue areas. The panels in the middle show hourly averages of radial velocity $u_r$, magnitude of $B_r$, number density, radial wave speed $v\!=\!u_r+v_A$ and the Alfv\'enic Mach number $M_A\!=\!u_r/v_A$. For the 12th encounter the 24h period prior 19:00UT May 31 is shown as there is no public plasma data after that. The wave speed varies along the PSP orbit in both cases. The figures on the right show the high cadence observations of the three components of the magnetic field for half hour periods. The background is removed by subtracting a 7.4 minute sliding average. During the first encounter all three components vary with similar amplitudes. For the second encounter dBR is multiplied by 4 to make its variation similar to the perpendicular oscillations. The dBR$-$dBT and dBR$-$dBN correlation coefficients r(dBR,dBT) and r(dBR,dBN) calculated over a two-minute sliding window show strong correlations of dBR with the tangential components. The light red rectangles highlight where dBR is highly correlated with dBT or dBN. The yellow rectangles highlight anti-correlation, where the signs are opposite. The green rectangles indicate times when $B_r$ is approximately constant. The bottom left panel shows one of these times. The solid line is $B_r\!-\!800\,$nT\,$\approx\!0$. The dotted and dashed lines show dBT and dBN varying with similar amplitudes consistent with roughly circularly polarized \alf waves.
}
\label{fig:observations}
\end{figure}

Magnetic switchbacks are rapid high amplitude reversals of the radial magnetic field in the solar wind that do not involve a heliospheric current sheet crossing. First seen sporadically in the seventies in Mariner and Helios data, switchbacks were later observed by the Ulysses spacecraft beyond 1 au and have been recently identified as a typical component of solar wind fluctuations in the inner heliosphere by the Parker Solar Probe spacecraft. Here we provide a simple yet predictive theory for the formation of these magnetic reversals: the switchbacks are produced by the shear of circularly polarized Alfv\'en waves by a transversely varying radial wave propagation velocity. We provide an analytic expression for the magnetic field variation, establish the necessary and sufficient conditions and show that the mechanism works in a realistic solar wind scenario.

The solar wind, to a good approximation, can be described with the equations of ideal magnetohydrodynamics (MHD). 
Parker Solar Probe observations of switchbacks show a tight correlation of magnetic and velocity perturbations that are characteristic of \alf waves \cite{Kasper:2019}.
\alf waves are typically thought of as transverse oscillations around a constant guide field $B_r$, which, in case of the solar wind, points approximately in the radial direction within Mercury's orbit.
The magnitudes of transverse velocity and magnetic perturbations, $u_{\perp}$ and $B_{\perp}$, are related as $u_{\perp} = B_{\perp}/\sqrt{\mu_0\rho}$, where $\rho$ is the mass density of the solar wind and $\mu_0$ is the magnetic permeability of vacuum. 
Circularly polarized \alf waves are in fact exact solutions of the MHD equations even when their amplitude $B_\perp$ is large.
The most puzzling property of the observed switchbacks is that the presumed guide field $B_r$
changes sign with frequent large amplitude oscillations.

This suggests that PSP observes 
spherically polarized \alf waves \cite{Barnes&Holl:1974}. For these waves both the magnetic field vector $\mathbf B$ and velocity vector $\mathbf u$ oscillate in arbitrary directions and
$\mathbf u = \pm \mathbf B/\sqrt{\mu_0\rho}$ in the coordinate frame moving with the wave, where the sign determines if the wave propagates parallel or anti-parallel with the magnetic field direction. 
This is fully consistent with PSP measurements \cite{Kasper:2019}. 

There is, however, a requirement for nonlinear spherical \alf waves to be an exact solution: the magnetic pressure $p_B=B^2/(2\mu_0)$ must be constant. 
This is actually not trivial, because a non-constant divergence-free magnetic field typically has a spatially varying amplitude with the exception of circularly polarized \alf waves \cite{Marris:1970}. 
Spherically polarized \alf waves are only approximately stationary, but they can travel large distances in the solar wind without significant dissipation. 
There have been several ideas put forward how switchbacks form, including magnetic reconnection \cite{Drake:2021}, Kelvin-Helmholtz instability \cite{Mozer:2020, Ruffolo:2020}, compressible turbulence \cite{Mallet:2021, Squire:2020}, and radial velocity shears and jets \cite{Landietal:2006,Schwadron:2021}, 
but none of these provide a fully self-consistent explanation for all observed properties.


\begin{figure}
\begin{center}
\includegraphics[width=0.58\textwidth]{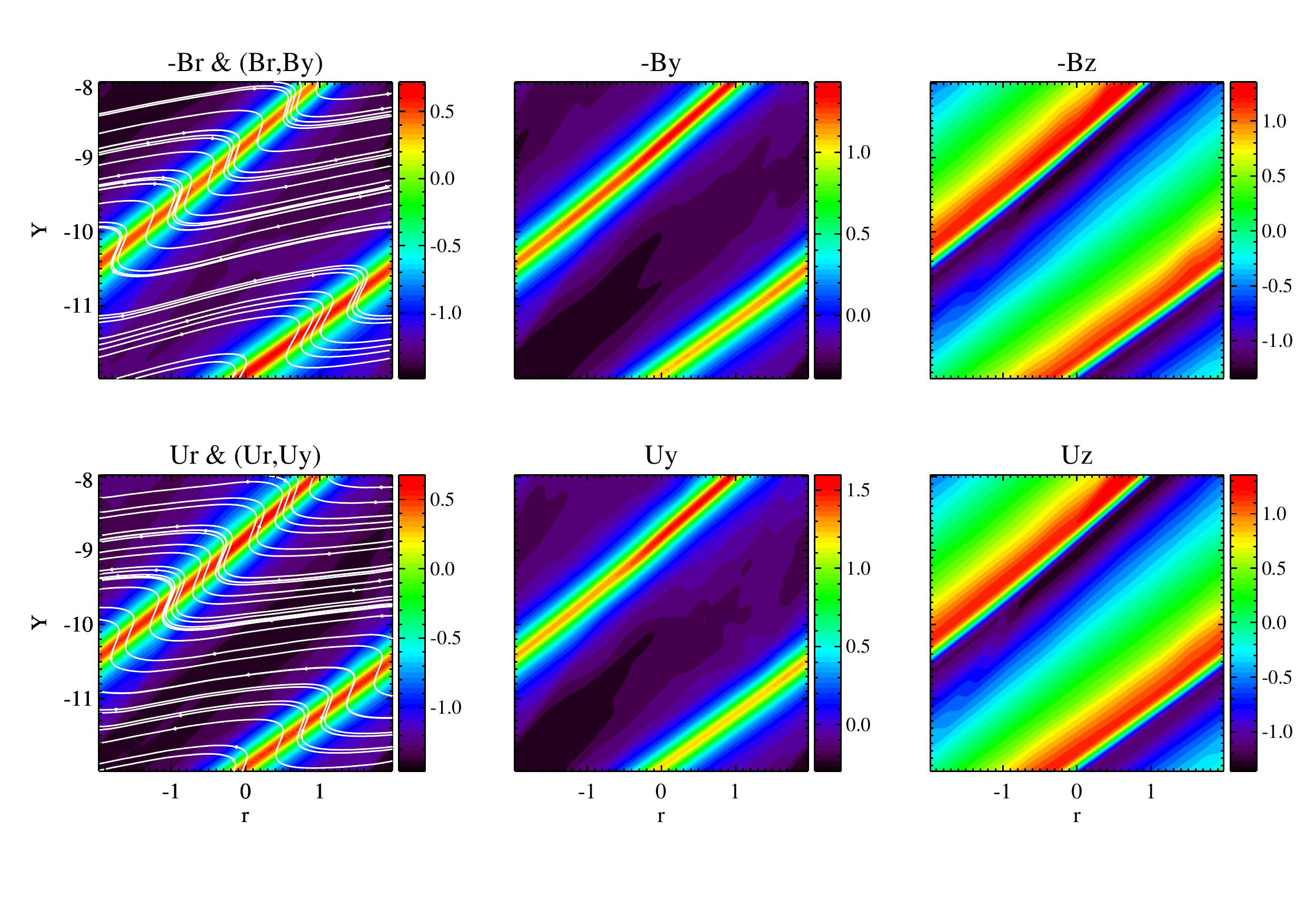}
\includegraphics[width=0.40\textwidth]{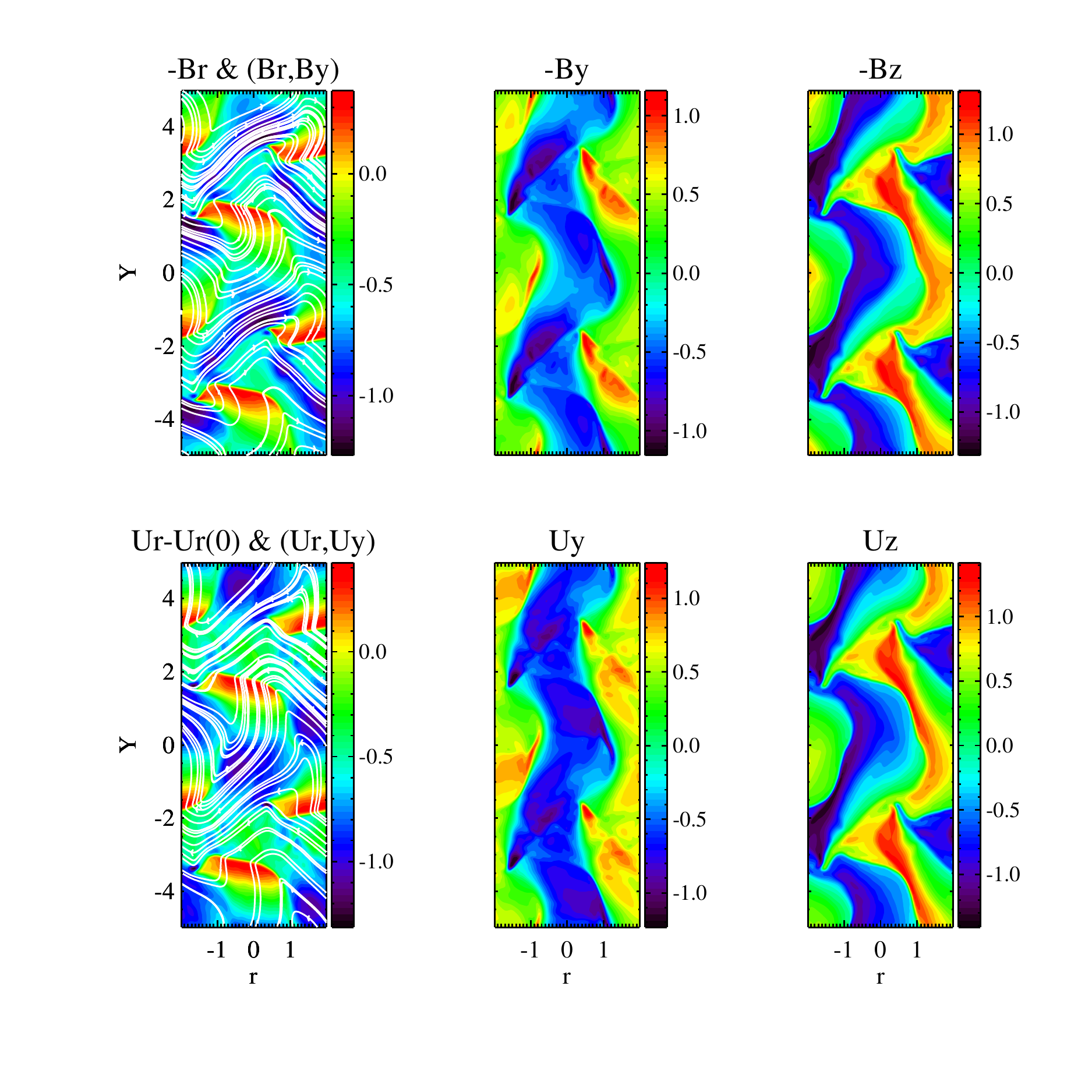}
 \end{center}
\vspace{-0.5cm}
\begin{center}
\ \ \ \ \includegraphics[height=0.3\textwidth]{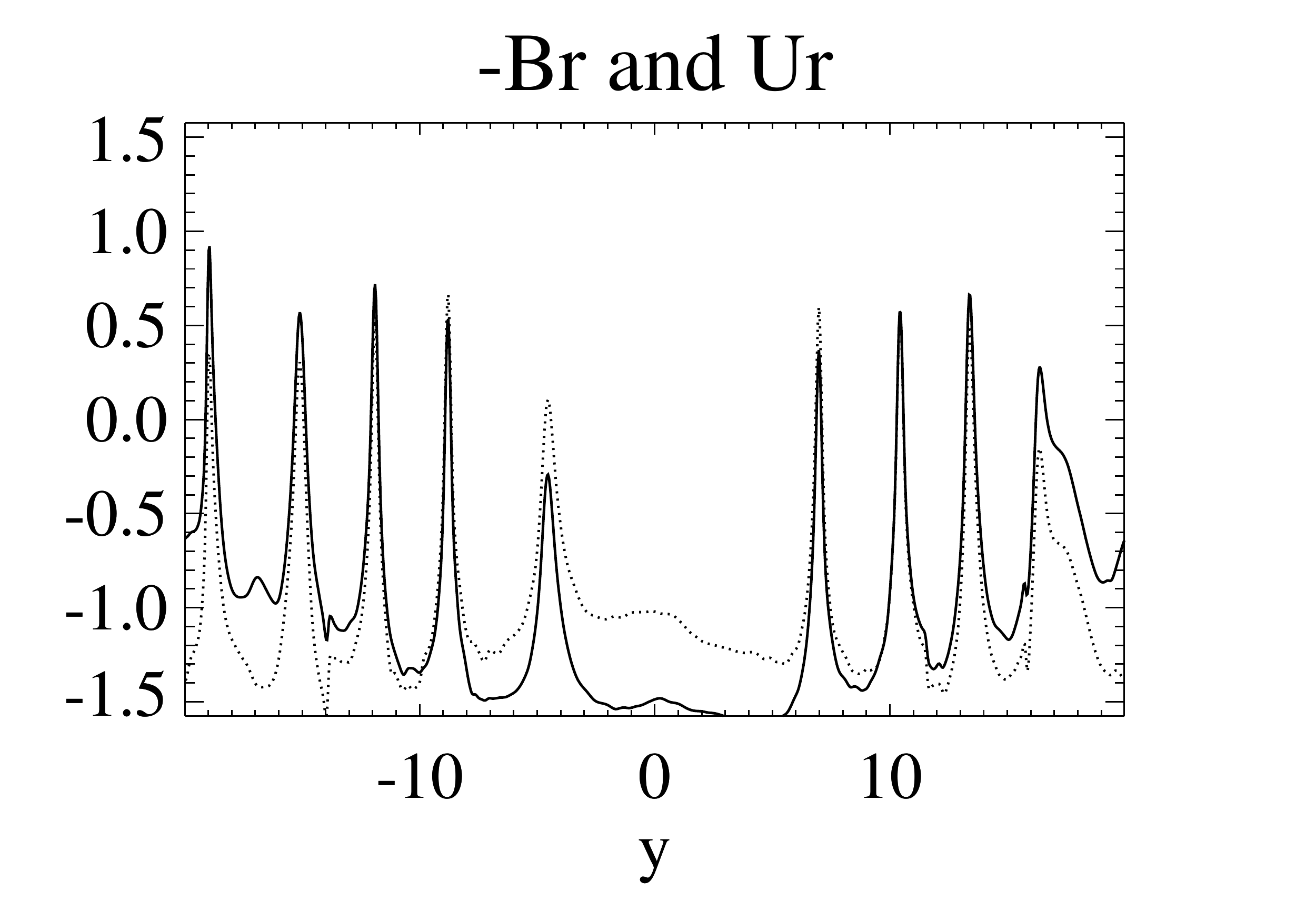}
\ \ \ \ \ \ 
\includegraphics[height=0.3\textwidth]{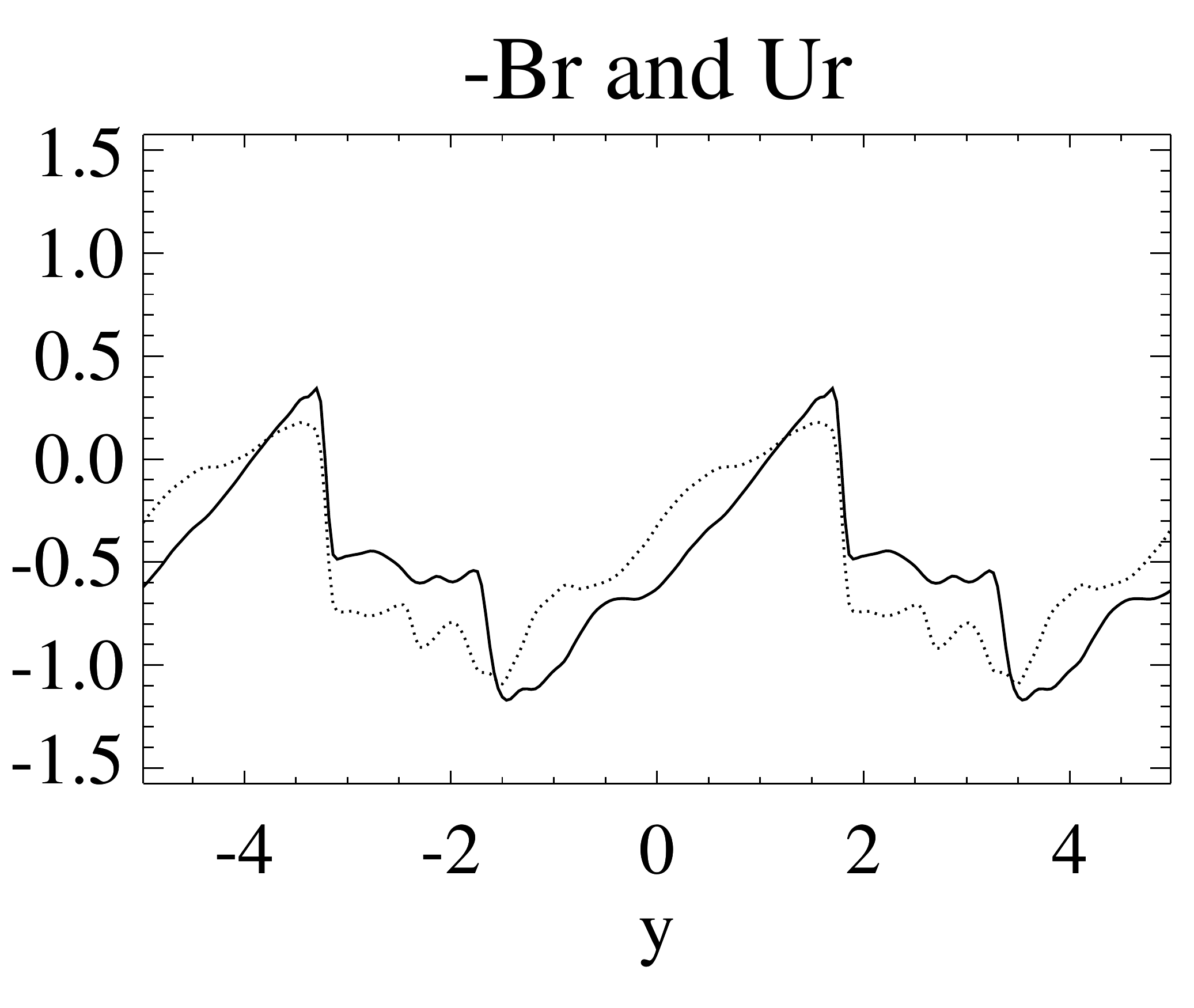}
\end{center}
\caption{Numerical solutions of sheared circularly polarized \alf waves in a double periodic box. Colors show components of the magnetic field and velocity. White lines are field lines and streamlines. Top left: 
The long wavelength $\lambda_y/\lambda_r=10$ shear produces a highly distorted \alf wave. Only a part of the domain is shown where the shear is maximal. Parameters:
$B_r=1$, $p=1$ and $v_1=0.5$ (by perturbing $B_r$) and time $t=20$.
Top right: The comparable wavelength shear $\lambda_y/\lambda_r=1.25$ results in a complex spherically polarized \alf wave solution. Parameters:
$B_r=0.5$, $p=1$, $v_1=0.5$ (by perturbing $u_r$) and $t=6.6$.
Bottom: cuts along $r=0$. Solid lines show the $-B_r$ magnetic field component, while the dotted lines are the $u_r$ velocity components coordinate system moving with the wave. 
\label{fig:sheared}}
\end{figure}

We propose a new explanation for the formation of switchbacks 
and provide supporting observational, theoretical and numerical evidence. \textbf{The switchbacks are produced by circularly polarized \alf waves distorted and twisted by a transverse shear of the radial wave speed.} 
The radial speed of an outward traveling \alf wave is $v\!=\!u_r+v_A$, where $v_A\!=\,\mid\!\!B_r\!\! \mid\!\!/\sqrt{\rho\mu_0}$ and $u_r$ are the \alf and solar wind speeds in the radial direction, respectively. 
The wave velocity can vary for three reasons: variation of $u_r$, $B_r$, or  $\rho$. Our numerical tests confirm that any of these can produce switchbacks.

Let us consider a sinusoidally sheared radial wave velocity profile $v(y)=v_0 + v_1\sin(2\pi y/\lambda_y)$, where $\lambda_y$ is the wavelength in the $y$ direction that is perpendicular to the radial direction and $z$ completes the coordinate system. 
The wave velocity shear impacts an initially circularly polarized sinusoidal \alf wave with radial wave length $\lambda_r$. 
The magnetic field lines of the wave oscillate within a width $w = (B_{\perp}/B_r)\lambda_r/\pi$. 
A long wave-length velocity perturbation, $\lambda_y \gg w$, will shear the circularly polarized waves and rotate the field in the $r$--$y$ plane across several waves.
The left side panels of Figure~\ref{fig:sheared} show numerical simulation results for this case. 
When $\lambda_y\sim w$, a much more complex solution emerges as shown in the right panel. 
Finally, for $\lambda_y\ll w$ the velocity shear can bend the transverse field lines as found by \cite{Landi:2005}, but this only works if the magnetic field is weak, which is not the case near PSP. 

\begin{figure}
\begin{center}
\includegraphics[width=\textwidth]{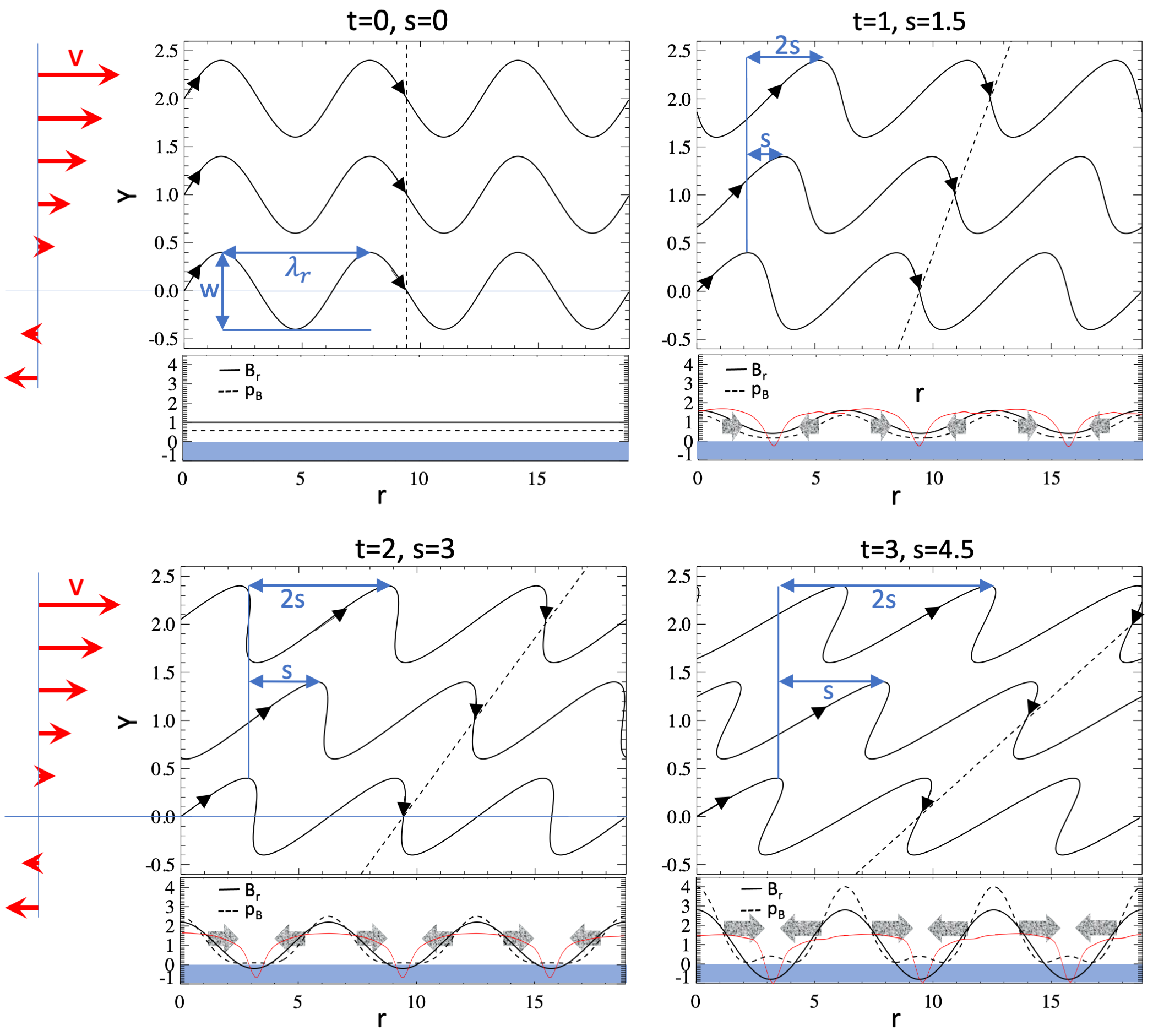}
\caption{Illustration how switchbacks form. The top part of the four panels show the evolution of three magnetic field lines (solid black curves with arrows) projected to the $r-y$ plane. The dashed lines follow $r=3\pi+sy$ indicating the amount of shear $s=1.5t$ that grows linearly with time $t$.
At time $t=0$ the original circularly polarized \alf wave is shown with wavelength $\lambda_r=2\pi$ and width $w=(B_\perp/B_r)\lambda_r/\pi=0.8$  At this time $B_r$ is uniform and the magnitude of the perpendicular field is $B_\perp=0.4 B_r$. The wave velocity $v=1.5y$ shown by the red arrows changes linearly with $y$. At time $t=1$ the field lines are mildly distorted. By time $t=2$ the field line folds over and $B_r$ changes sign creating a switchback around the positions where the dashed line intersects the field lines. At $t=3$ there is substantial radial field reversal. The bottom part of the panels show $B_r(t)=B_r(t=0)+sB_y$ (solid lines) and the magnetic pressure 
$p_B=(B_\perp^2+B_r^2)/(2\mu_0)$ (dashed lines) at the four time instances. Switchbacks occur when the line enters the blue regions. The gray arrows show the gradient of the magnetic pressure that compresses the plasma and the magnetic field. The red lines show numerical simulation results for comparable conditions. The switchbacks are narrow peaks between flatter background field regions.  }
\label{fig:theory}
\end{center}
\end{figure}

The long wavelength case can be regarded locally as a constant shear of the radial wave velocity, $dv/dy=\mathrm{const}$, which
can be studied analytically. 
Let us consider a circularly polarized wave with $B_r=\mathrm{const}$, $B_y=B_{\perp}\cos r$ and $B_z=B_{\perp}\sin r$, 
so the wavelength is $\lambda_r=2\pi$.
After time $t$, the field lines at a distance $y$ from the center of the wave will be pushed from position $r$ to $r'=r+sy$, where $s=t(dv/dy)$ is the shear at time $t$.
To a first order approximation, the shear will simply shift $B_y$ and $B_z$ in the radial direction:
$B'_y(r,y)=B_y(r-sy)$ and $B'_z(r,y)=B_z(r-sy)$ as illustrated in Figure~\ref{fig:theory}.
On the other hand, the originally constant $B_r$ will change to $B'_r(r,y)=B'_r(r-sy)=B_r + s B_y(r-sy)$ that varies proportionally to $B_y$.
A switchback occurs when $B'_r$ changes sign. This happens when the shear
exceeds the ratio of the radial and transverse field magnitudes in the original circularly polarized wave: $s>B_r/B_{\perp}$. The observations shown in 
Figure~\ref{fig:observations} support this assertion too: there are several switchbacks during encounter 1 when the average $B_r\approx -63\,$nT with oscillation amplitudes dBR$\,\approx 32\,$nT, and the average $B_\perp\approx\sqrt{\mathrm{dBT}^2+\mathrm{dBN}^2}\approx 49\,$nT suggesting $s\approx 0.65$, which is comparable to $B_r/B_\perp\approx 1.3$. 
During encounter 12, $B_r$ varies from 400\,nT to 800\,nT and the average  $B_\perp\approx 210\,$nT is about four times larger than dBR implying $s\approx 0.25 \ll B_r/B_{\perp}\approx 2$ to 4, so there are only a few switchbacks. 
The observations also show strong correlations between the oscillations of  $B_r$ and the perpendicular components at most times. This confirms that the oscillations are the radial and perpendicular components of a sheared oscillation. The direction of the shear determines if dBR is proportional to dBT or dBN (or some linear combination of them), and the sign of $s$ determines if there is a positive correlation or an anti-correlation. 
On the other hand, the ratio of amplitudes is fairly constant for each encounter suggesting that the average shear is a function of radial distance from the Sun, or in other words, it is increasing in time as the wave propagates outward.

The first order approximation satisfies the divergence-free property, but the magnetic pressure $p'_B=(B^2+2sB_r B_y' + s^2B'^2_y)/(2\mu_0)$ is no longer constant. The magnetic pressure gradient will compress the plasma and modify $B_x'$ and $B'_y$ while maintaining the $B'_r(r-sy)=B_r+sB'_y(r-sy)$ relationship so that the field remains divergence free. 
The plasma will move towards the small magnetic pressure region where $B_r$ is small and the switchbacks form. This explains why the observed switchbacks are narrow peaks while the regions with normal $B_r$ direction are wide and flat (see Figure~\ref{fig:observations}). 

An additional requirement for a switchback to occur is that the shear velocity has sufficient energy to distort the original wave.
A simple estimate is that the average energy density of the shear motion is comparable to, or larger than, the magnetic energy density of the transverse magnetic field: $\rho v_1^2 > C(B_{\perp})^2/\mu_0$, where $C\approx0.2$ based on numerical experiments.

Finally, the turning needs to happen fast enough while the wave is traveling outward in the solar wind. It takes the waves $t=D/\bar v$ to reach the spacecraft, where $D$ is the distance from  the location where the circularly polarized \alf waves start to get sheared, and $\bar v$ is a proper average of the radial wave velocity. 
For a switchback to occur, 
$s=(D/\bar v)(dv/dy)>B_r/B_\perp$ is required. 

The shearing does not continue indefinitely. Eventually the energy related to the shear is exhausted and the perturbed waves will keep propagating with minimal evolution. The hourly averaged PSP plasma data during the first encounter (see Figure~\ref{fig:observations}) suggests that this is indeed happening. $U_r$, $B_r$, and $\sqrt{\mu_0\rho}$ vary 5\%, 29\% and 6\%, respectively, which would result in $\approx\!30\%$ variation in the wave speed $v$ if these were independent of each other. But the observed wave speed only varies 5.5\%, which means that the velocity, magnetic field and density variations contributing to the wave speed cancel each other out. This cancellation is caused by the distortion of the field reducing the energy of the shear as the system tries to find an approximate equilibrium solution with a constant wave speed. If the energy density of the shear exceeds the magnetic energy density by orders of magnitudes, then the shear will eventually break down due to non-ideal MHD processes, such as magnetic reconnection or turbulent cascade to kinetic scales. 


The basic dynamics of shearing a circularly polarized \alf wave can be captured in a two-dimensional (2D) MHD simulation with three vector components for velocity and magnetic field. 
The simulation domain is a double periodic rectangle. 
The $r$ direction corresponds to the radial direction in the solar wind. The frame of reference is chosen such that the initial circularly polarized wave, without the perturbation of the wave speed, is at rest. The setup is normalized by setting the units of distance, time and mass, so that $\lambda_r=4$,  $B_{\perp}=u_{\perp}=1$, and $\mu_0\bar\rho=1$ where the $\bar\rho$ is the unperturbed density.
The initial magnetic and velocity fields are $B_y=-u_y=\cos(2\pi r/\lambda_r)$ and $B_z=-u_z=-\sin(2\pi r/\lambda_r)$, which correspond to the \alf wave propagating in the $+R$ direction relative to the plasma.
There are only four free dimensionless parameters: the relative strength of the unperturbed guide field $\bar B_r/B_{\perp}$ (which also determines  $\bar u_r=-\bar v_A = -\bar B_r$ to make the wave standing), the plasma beta $\bar \beta=p/\bar p_B$ that defines the pressure $p$, and the two parameters, $v_1/u_{\perp}$ and $\lambda_y/\lambda_r$, for the wave velocity perturbation $v_1\sin(2\pi y/\lambda_y)$. 
We can perturb either $u_r$, $B_r$ or $\rho$ to change the wave speed.
The size of the domain in the $r$ direction is $\lambda_r$, while  in the $y$ direction a multiple of $\lambda_y$. 

The simulations are performed with the BATS-R-US code \cite{Powell:1999, Toth:2012swmf} on a fine grid (cell size $\Delta r=\Delta y=0.04=\lambda_r/100$) with a fifth order accurate scheme \cite{Chen:2016}.
The left panels of Figure~\ref{fig:sheared} show the solution for the long wavelength case, with the perturbation applied to $B_r$, in the part of the domain where the shear is near maximal. The result is a distorted wave, similar to the analytic description, with large switchbacks (left bottom panel) that look remarkably similar to the observations in Figure~\ref{fig:observations}.
The right panels show the solution for a case when the wave length of the perturbation $\lambda_y$ is comparable to  $w$.
The solution shows complex structures that do not resemble a circularly polarized wave, still the Alfv\'enic relations, $-B_y\approx u_y$, $-B_z\approx u_z$ and $-B_r\approx u_r-v_1\sin(2\pi y/\lambda_y)$, hold (subtracting the initial perturbation from $u_r$ removes the background variation). In this case the $y=0$ cuts show more complicated switchback structures. 

\begin{figure}
\begin{center}
\includegraphics[width=\textwidth]{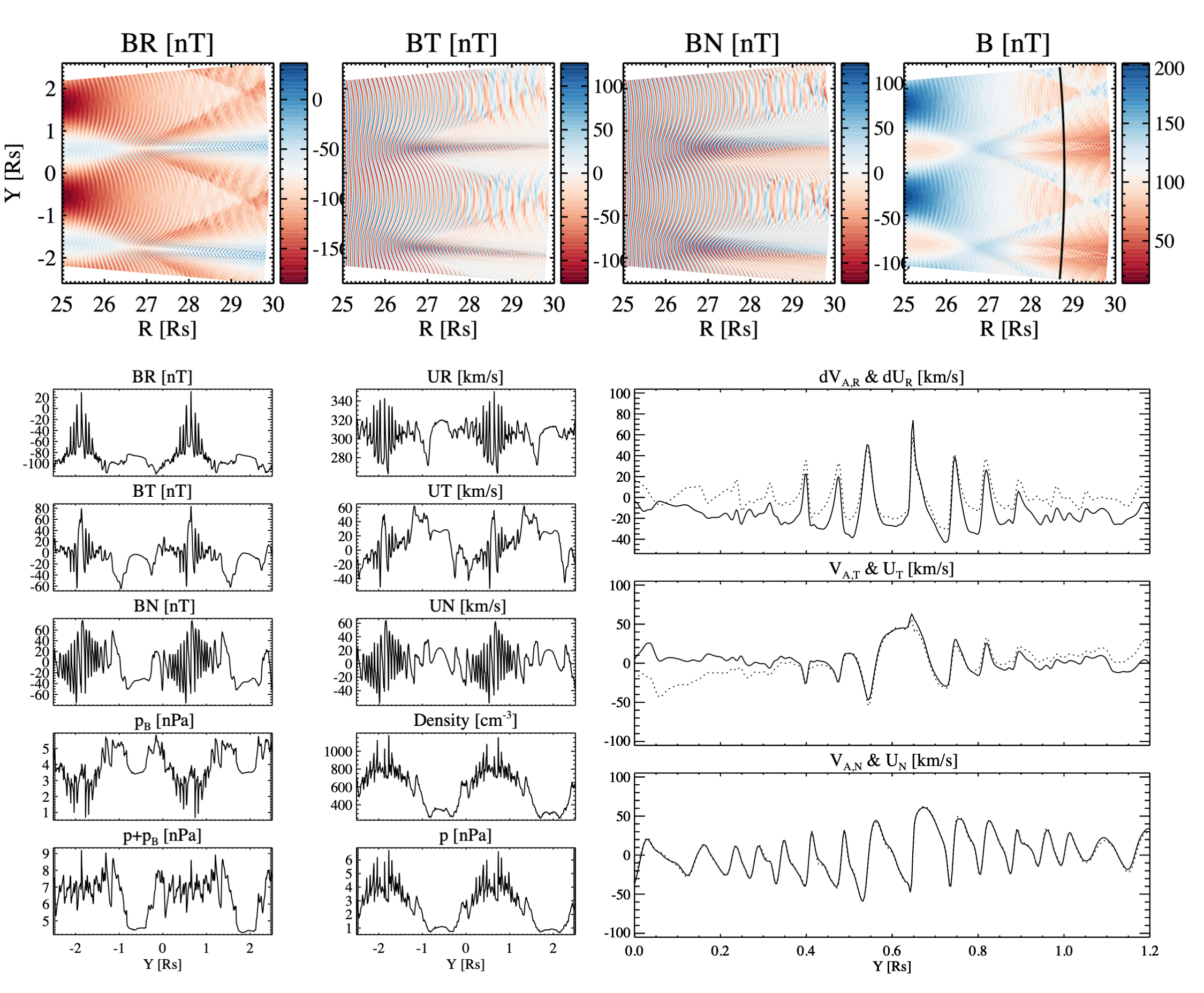}
\caption{
Formation and propagation of spherically polarized \alf waves in the spherically expanding solar wind. The top panel shows the three components of the magnetic field and its magnitude in part of the 2D computational domain. The circularly polarized \alf waves enter through the left boundary at $R=25\,R_s$. The incoming radial field is perturbed along the $Y$ direction, which causes a shear in the \alf wave speed and the development of spherical polarization. Switchbacks with $B_r>0$ form at $r>27.5\,R_s$. The total magnetic field (top right) is relatively smooth. The black curve indicates a possible PSP trajectory at $r\approx 29\,R_s$. The bottom left panel shows the magnetic field and velocity components as well as the magnetic pressure $p_B$, the density, the thermal pressure $p$, and the total pressure $p+p_B$ along the trajectory. All quantities are comparable to PSP observations during the first encounter. The gradients of the total pressure are small, but not zero. Density variations are also substantial. The bottom right panel compares the magnetic (solid lines) and velocity (dotted lines) perturbations around a switchback. The magnetic field components are converted to \alf velocity components: $\mathbf{V}_A = \mathbf{B}/\sqrt{\rho\mu_0}$. For the radial components the background variation is removed with a smoothing over 100 grid cells ($0.28\,R_s$). All components satisfy the Alfv\'enic relationship to a high accuracy similar to PSP observations \cite{Kasper:2019}. 
}
\label{fig:solarwind}
\end{center}
\end{figure}
Finally, we show that the mechanism also works in the radially expanding solar wind. We use physical units for easier comparison with observations.  The 2D computational domain is a spherical wedge extending from $r=25\,R_s$ to $40\,R_s$ and the azimuthal angle goes from $-5^\circ$ to $5^\circ$.
The 2D computational grid consists of $4,000\times1,600$ cells. The boundaries are periodic in the azimuthal direction and outflow condition is applied at $r=40\,R_s$.
The circularly polarized \alf waves enter at $r=25\,R_s$ with amplitude $B_\perp=80\,$nT and wavelength $\lambda_r=0.1\,R_s$. 
The number density, the radial velocity and the temperature are $800\,$cm$^{-3}$, 300\,km/s, and 350,000\,K, respectively.
The radial field is $B_r=-120+64.8\sin(2\pi y/\lambda_y)\,$nT and $\lambda_y=2.18\,R_s$, which is half of the width of the domain at the inflow boundary.

Figure~\ref{fig:solarwind} shows the solution at $t=10$ hours, which is enough for the solar wind to propagate from $25\,R_s$ to $40\,R_s$ with 300\,km/s speed. The figures shows that switchbacks develop with their characteristic asymmetric shapes and the Alfv\'enic relationship between magnetic and velocity fields are satisfied. This simulation was set up to demonstrate the formation of switchbacks in an idealized solar wind. The real solar wind is 3-dimensional with a spectrum of \alf waves that become turbulent due to the spherical expansion \cite{Dongetal:2014}.
According to previous theoretical and numerical studies \cite{Squire:2020, Mallet:2021} the turbulence will preserve the spherically polarized \alf waves and further enhance their amplitudes. 

This paper focused on explaining the puzzling observations by PSP, but the interaction of wave velocity shear with circularly polarized \alf waves 
can play an important role in 
the physics of the solar wind. The interaction 
can create mode conversion from \alf turbulence to compressive turbulence
heating and accelerating the solar wind \cite{Akhavan:2022}. 

\bmhead{Acknowledgments}

G. T\'oth and B. van der Holst are supported by NSF grant PHY-2027555 and NASA grant 80NSSC22K0892. PSP data was obtained through NASA CDAWeb. Simulations were performed on the Pleiades supercomputer at NASA Ames. BATSRUS is open source at http://github.com/MSTEM-QUDA. We thank Prof. Tamas Gombosi at the University of Michigan for excellent comments and suggestions.






\end{document}